# Experimental evidence on the Altshuler-Aronov-Spivak interference of the topological surface states in the exfoliated Bi$_2$Te$_3$ nanoflakes

Zhaoguo Li (李兆国)[1], Yuyuan Qin (秦宇远)[1], Fengqi Song (宋凤麒)[1,2 *],

Qiang-Hua Wang (王强华)[1], Xuefeng Wang (王学锋)[1], Baigeng Wang (王伯根)[1],

Haifeng Ding (丁海峰)[1], Chris Van Haesondonck[2], Jianguo Wan (万建国)[1], Yuheng

Zhang (张裕恒)[3], Guanghou Wang (王广厚)[1]

1 National Laboratory of Solid State Microstructures, Nanjing University, Nanjing, 210093, P. R. China

2. Laboratorium voor Vaste-Stoffysica en Magnetisme, Kathelieke Universteit Leuven, B3001, Belgium

3. High Magnetic Field Laboratory, University of Science and Technology of China, Hefei, 230027, China

**Abstract:** Here we demonstrate the Altshuler-Aronov-Spivak (AAS) interference of the topological surface states on the exfoliated Bi$_2$Te$_3$ microflakes by a flux period of h/2e in their magnetoresistance oscillations and its weak field character. Both the osillations with the period of h/e and h/2e are observed. The h/2e-period AAS oscillation gradually dominates with increasing the sample widths and the temperatures. This reveals the transition of the Dirac Fermions' transport to the diffusive regime.

---

[*] Corresponding authors. Email: songfengqi@nju.edu.cn. Fax: +86-25-83595535.

The Aharonov-Bohm (AB) interference [1] has been recognized as one of the most intuitive features to identify the surface state (SS) in the current field of topological insulators (TI) [1-4,5-6]. The SS, exhibiting massless Dirac Fermions with an intrinsic spin helicity, is determined by the abrupt evolution between two topologically distinct gapped spaces (e. g. vacuum and the TI material). The backscattering of the Dirac carriers by impurities are forbidden, hence the surface carriers are expected to be immune from impurities. Such topological SSs are therefore aspired for applications in spintronic devices and quantum information technologies. On a TI flake with a piercing magnetic field (B), the SS will enclose a magnetic flux. Traditional theory shows two periods of the magnetic flux, $\Phi_0$ (=h/e, the flux quantum) and $\Phi_0/2$, on the magnetoresistance (MR) oscillations [7]. For a practical sample, we check the product flux of the oscillating period of B ($\Delta B$) and its cross section ($S_{TI}$). If $\Delta B \cdot S_{TI}$ equals to the integral times of $\Phi_0/2$, we will be convinced of a conductive SS in the TI flake [1,5,8]. The anomalous Berry phase leads to a phase shift of $\pi$ in the AB MR oscillations[7]. Such MR oscillations have been observed in the nanoribbons of $Bi_2Se_3$ and $Bi_2Te_3$ [1-3]. The quantum dephasing and mean free paths of the Dirac carriers can be then discussed [1,3,7,9].

The MR oscillations with the period of $\Phi_0/2$ deal with a different physical picture to the h/e-periodic ones besides the identification of the SSs. It is the Altshuler-Aronov-Spivak (AAS) quantum interference due to the weak (anti)localizations. Firstly, in an AAS picture, the MR period of $\Phi_0/2$ is obtained by the interference of a pair of time-reversal loops. The electrons collect the same phases while travelling by the scattering centers along the two reversed loops, therefore, the MR contributed by an AAS picture will survive the environment of strong disorders within the space of phase coherence[10]. Secondly, the spin-helical electrons in the two loops have opposite spins. It was argued that the AAS interference would be suppressed in the case of ideal helical carriers. [1-2] The study of the AAS interference will then probe the decay of the spin helicity. Thirdly, Dirac carriers have been demonstrated to exhibit rather long spin relaxation and dephasing lengths in graphene.[11] The h/2e-period AAS oscillation is predicted to be dominant in the case of strong disorders when the Fermi level is away from the Dirac point. [7] It is solely affected by the phase coherence, which is an important factor in the quantum transport. However, the AAS interference of the TI's SS have not been highlighted. In this Letter, we report the experimental evidence on the AAS quantum interference of the topological SS on the exfoliated $Bi_2Te_3$ microflakes. Its dependence on the samples' dimension and temperature are also observed.

The samples of $Bi_2Te_3$ TI flakes were prepared on a silica-capped silicon wafer following a typical exfoliation procedure similar to the work of graphene.[12-13] The polycrystalline $Bi_2Te_3$ clumps were obtained from Alfa Aesar (vacuum deposition grade). The ribbons with uniform widths were selected. After annealed in a Te atmosphere, they were checked by the Raman spectroscopy, where three clear "g" peaks at about 60, 103 and 134 $cm^{-1}$ stand for good spatial orders of the measured samples. [12] The two-probe electrodes were then applied onto the flakes by a standard lift-off procedure, indicating only the change of the resistance is reliable. [14] All the MR measurements were carried out in the Quantum Design Physical Property Measurement System (PPMS)-9 and PPMS-16 systems, when the magnetic field was aligned parallel to the current. The errors with the resistance measurements were less than 0.05 Ω. There was a large risk of failure with exfoliating the inuniform bulk, microfabrication and measurement damage. Furthermore, the temperature is not lower than 2K. We therefore only measured 5 samples. Fast Forier transformations (FFT) were carried out.

Two flux periods of $\Phi_0$ and $\Phi_0/2$ are generally expected, and have been indeed observed in the MR oscillations of our $Bi_2Te_3$ microflakes. **Figure 1(a)** shows an atomic force microscopic (AFM) image of a $Bi_2Te_3$ nanoribbon, the height profile of which in the inset reveals a cross section of about 0.9 μm × 63 nm ($S_1$). Its MR curve with an oscillating period ($B_{p1}$) of ~700 Gauss is displayed in Fig. 1(b). Apparently, $S_1 \cdot B_{p1}$ equals $\Phi_0$ within its tolerance. The power spectrum from an FFT of the data, shown as the inset of Fig 1(b), reveals two clear periods , i.e. $\Phi_0$ and $\Phi_0/2$ [15]. (We also compare the MR curves to those of a gold ribbon measured in the same runs but find no apparently periodic MR oscillation.) The periodic MR oscillations confirm the presence of a conductive SS[1].

The $\Phi_0/2$ period becomes dominant in MR curves of some large sheets. **Figure 2**(a) and (b) show the AFM image and MR curve of a broader microflake, respectively. The cross section of the sample is $S_2$= 4.8 μm × 73 nm, and the magnetic field period of the MR is $B_{p2}$ = 54 Gauss. The FFT spectrum in the inset shows a dominant flux period of $S_2 \cdot B_{p2} \approx \Phi_0/2$. Obviously, the carriers traveled across different crystalline planes and different interfaces (TI-vacuum and TI-wafer) in a single loop. Hence, we conclude that the MR oscillations are induced by the topological SS rather than a conductive SS confined in a specific crystalline plane.[16] One may find some roughness in the present samples, which casts a confusion that such edges may terminate the phase coherence of the surface carriers and destroy the AB or AAS interference. However, note that all the samples that have been observed with the AB interference are nanoribbons of rectangular cross sections and sharp edges [1,3]. Therefore, the robust quantum interference against the roughness is attributed to scattering-free transport of topological SSs.

To identify the AAS origin of the $\Phi_0/2$ oscillation, several analyses have been performed. Firstly, the ratio between the amplitudes of the $\Phi_0/2$ and $\Phi_0$ oscillations (defined as the $\Phi_0/2$: $\Phi_0$ ratio) is checked as the temperature increases from 2 K to 64 K as shown in **Figure 3(b)**. A steady predominance of the $\Phi_0/2$ oscillations over the $\Phi_0$ oscillations has been found although both amplitudes decrease with increasing temperatures. The $\Phi_0/2$ oscillation stays trackable and dominant at 64 K, as seen in the lower panel of Fig 1(c). The thermal stability akin to the $\Phi_0/2$ oscillation can only be reasonably ascribed to the AAS effect, a fact to which we will return. Secondly, besides the MR flux period of $\Phi_0/2$, the other critical character of the AAS interference is its weak-field character due to its origin of weak localizations.[17] In the present experiment, the $\Phi_0/2$ period of the MR oscillation is clearly resolvable under weak fields in the range of 0-0.5 T, but is invisible when a medium magnetic field of 0.5-1.0 T is applied (Fig 1(c)). This is a signature of weak localization in the $\Phi_0/2$ oscillation. Finally, to pin down the AAS mechanism, we have to rule out possible $\Phi_0/2$ period from the so-called averaged AB oscillation at a finite temperature.[6,17-18] While the Fermi level $\mu_F$ is tuned away from the Dirac point, the $\Phi_0$-periodic MR curve evolves sensitively [7,9]. A small change of $\mu_F$ may result in a $\pi$ phase shift in the MR curve. The effect of a finite temperature is to average the zero-temperature MR curves around a central Fermi level. A superficial $\Phi_0/2$ period would appear by averaging the phase-shifted curves. However, such $\Phi_0/2$ oscillations should always follow that of the $\Phi_0$ period, and is expected to survive at higher magnetic fields. [1,17] These claims are inconsistent with the thermal stability and weak magnetic field requirement of the experimental $\Phi_0/2$ oscillations. We are therefore convinced that they are manifestations of the AAS quantum interference.

As a discussion, Fig. 3(a) schematically illustrates of the picture of AB and AAS interference in a TI nanoribbon. The AB interference is based on the phase collection in a single turn when the SS electron is threading around the TI bulk. This leads to the MR oscillation with the period of $\Phi_0$. Obviously, the involvement of even a single defect will introduce an arbitary phase to the SS electrons, which will smear the AB MR oscillations. The $\Phi_0/2$-period MR oscillation originates from two turns of the SS moving. Among all the possible interference paths, the pair of time-reversal loops will survive the defect scattering due to the same phase collection as shown in Fig 3(a), essentially the AAS interference. Hence, the AB oscillations are sample specific and suppressed by thermal smearing, while the AAS interference is solely suppressed by dephasing. The first issue is a bigger value for the dephasing length. The original work of Peng et al found that the AAS oscillation was suppressed[1]. A dephasing length of 0.5μm was obtained since the $\Phi_0$-period AB oscillation disappeared in the larger ribbons. In the present experiment, the AAS interference is demonstrated in some wider samples. Therefore, the dephasing lengths of a few micrometers can be reached. The large dephasing lengths in the exfoliated TI flakes confirm good crystalline states in our samples. This is consistent with the case of graphene, where the mechanically-exfoliated sheets maintain one of the best records on the electronic quality despite of some luck issues on the graphene preparations.[19]

Only the ratios between the amplitudes of $\Phi_0$ and $\Phi_0/2$ oscillations (the $\Phi_0 : \Phi_0/2$ ratio) is studied since a 2-probe configuration is employed. We can see the ratio decreases with the increasing of the temperatures and sample widths. Fig. 3(b) shows the temperature-dependent $\Phi_0 : \Phi_0/2$ ratio, measured in a sample with the width of 0.7μm. The $\Phi_0$ oscillation is dominant at 2 K, confirming the observation of Peng et al.[1] As the temperature increases, the $\Phi_0/2$ oscillation shows up and even becomes dominant at higher temperatures. The ratio reaches 1 at 8K. Fig.3(c) presents the size dependence of the $\Phi_0: \Phi_0/2$ ratio at a fixed temperature of 2 K. The $\Phi_0$ component is larger than the $\Phi_0/2$ one when width is smaller than 2 μm. The $\Phi_0/2$ component became dominant for larger flakes. The $\Phi_0/2 : \Phi_0$ ratio keeps increasing with increasing width. It reaches 1 at 1.5μm. We have also deposited some cobalt clusters on the samples and followed the change of the $\Phi_0/2 : \Phi_0$ ratio, which increases with increasing the Cobalt cluster coverage[2]. Based on the above descriptions on the $\Phi_0/2 : \Phi_0$ ratio, we suggest the AAS interference is more robust than the AB interference against the external perturbations such as increasing temperature, sample size and surface decoration. The temperature dependence can be attributed to the thermal smearing of the AB interference. Both increasing the sample width and cluster decoration introduce more opportunities of defect scatterings in a single AB interference path. Hence, the size dependence of the $\Phi_0 : \Phi_0/2$ ratio is in agreement with the crossover from a ballistic to diffusive transport of the TI's SS carriers.

**Conclusion**

We have demonstrated the AAS interference of the topological SS in the $Bi_2Te_3$ flakes by the MR oscillation with a period $\Phi_0/2$. The micrometered dephasing length are obtained in our samples at 2K. The AAS interferences are more robust than the AB interference against the increasing temperature, sample size and surface decoration. This may provide technical basis for potential applications of TI's in the quantum microdevices.

**Acknowledgements**

This work was financially supported by the National Key Projects for Basic Research of China (Grant No. 2010CB923401, 2011CBA00108 and 2009CB930501), the National Natural Science Foundation of China (Grant Nos. Grant Nos. 11023002, 11075076, 10904100, 61176088, 11134005, 51171077, 10974086 and 60825402), the PAPD project and the Fundamental Research Funds for the Central Universities. The first two authors (Zhaoguo Li and Yuyuan Qin) contribute equally.

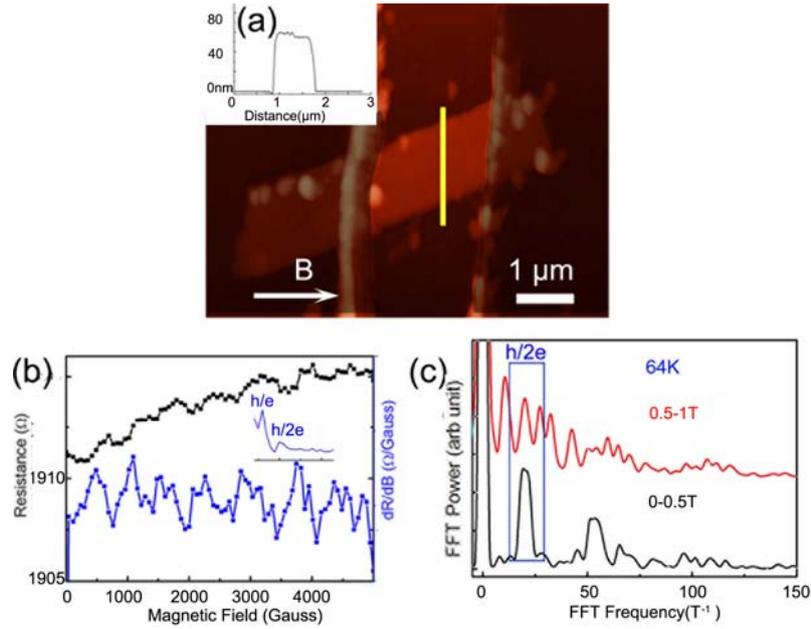

Figure 1. The MR oscillation of a $Bi_2Te_3$ flake: the $\Phi_0/2$ period and its weak-field character at 2K.  (a) An AFM image of the flake. The inset is the profile along the marked line, reading a cross section of 0.9μm*63nm ($S_1$). The electrodes are shadowed for eye guiding. (b) MR curve (black) and its field derivative (blue). the inset is its FFT power spectrum. Note that the nanoribbon is 15 degrees away from the field, which leads to a minor correction of cos 15°(=0.97).  (c): The change of the FFT spectra of the MR curves at 64K when the magnetic field is increased from 0-0.5T to 0.5-1.0T. The peak at 55$T^{-1}$ is attributed to the high-frequency FFT harmonics.

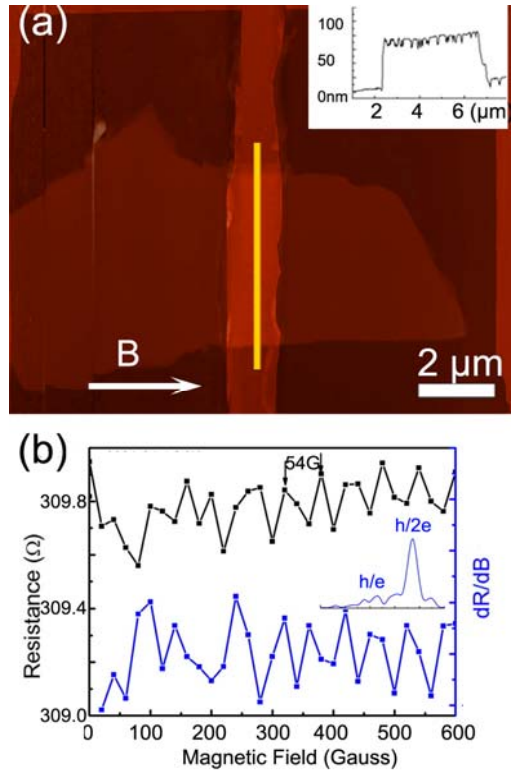

Figure 2. The MR oscillation of a broad $Bi_2Te_3$ flake (a) The AFM image. The inset is the height profile along the marked line. It reads a cross section of $S_2$= 4.8μm*73nm. The electrodes are shadowed for eye guiding. (b) MR curve (black) and its field derivative (blue). A dominant $\Phi_0/2$ peak is found in the FFT spectrum shown in its inset.

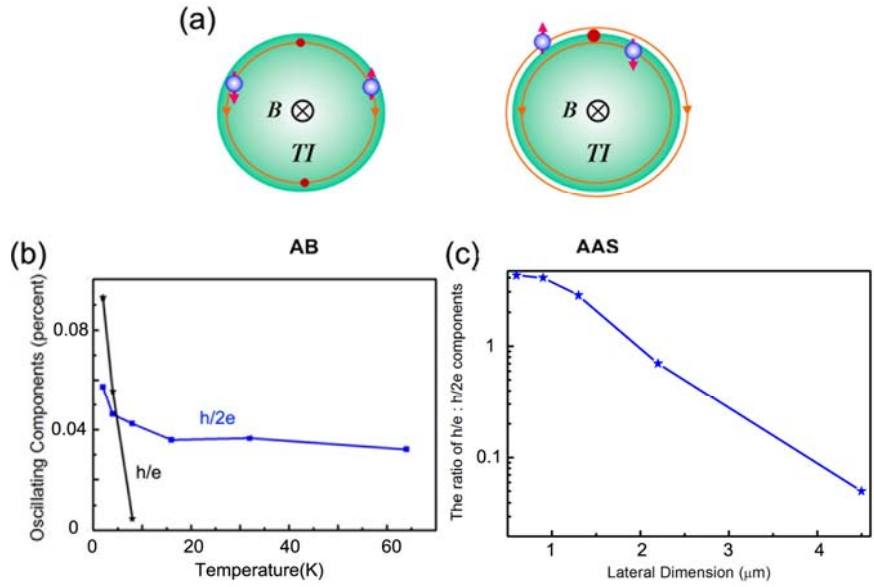

Figure 3 The MR oscillations with h/e and h/2e periods and their dependence. (a) The illustrations of a typical AB and AAS procedure. (b) The temperature-dependence of the $\Phi_0$: $\Phi_0/2$ ratio. (c) The dependence of the $\Phi_0$: $\Phi_0/2$ ratio on the sample width at 2K.